\documentclass{PoS}

\title{
Spontaneous mass generation suggests small dimension of the SM group $U(1)\times 
SU(2)\times SU(3)$
}

\ShortTitle{Small dimensions of $U(1)$, $SU(2)$, $SU(3)$ and fermion masses }

\author{\speaker{Felipe J. Llanes-Estrada}\thanks{Work supported by grants from MINECO FPA2011-27853-C02-01 and FPA2016-75654-C2-1-P, and carried out in the inspiring atmosphere of the theoretical physics department and UPARCOS.}\\
        E-mail: \email{fllanes@fis.ucm.es}}

\author{Guillermo Garc\'{\i}a Fern\'andez and Jes\'us Guerrero Rojas\\
        Departamento de Fisica Teorica I, Plaza de las Ciencias 1, Fac. CC. Fisicas; Universidad Complutense de Madrid, 28040 Madrid, Spain\\
        }

\abstract{
The reasons behind the gauge symmetry of the Standard Model, $U(1)\times SU(2)\times SU(3)$, are still unsettled. One obvious feature is the low dimensionality of all its subgroups. 
Under certain conditions, a negative answer to the question \emph{why not larger groups like SU(15), or for that matter, SP(26) or E7?} is possible. \vspace{0.4cm}

 We have recently observed that fermions charged under large groups acquire much bigger dynamical masses, all things being equal at a high e.g. GUT scale, than ordinary quarks. Should such multicharged fermions exist, they are just too heavy to be observed today 
(and have either decayed early on if coupled to the rest of the Standard Model, or become reliquial dark matter if uncoupled).
Their mass scale is dictated by strong antiscreening of the running coupling for those larger groups (with an appropriately small number of flavors) together with scaling properties of the Dyson--Schwinger equation for the fermion mass.\vspace{0.4cm}

The generated fermion mass (assuming only few flavors, to avoid spoiling antiscreening) grows exponentially with the number of colors as 
$M(N_c) \propto e^{N_c} \times \theta(N_f^{\rm critical} -N_f)$
for scales much below the GUT scale. Large groups would be strongly coupled already near the GUT scale and fermions charged thereunder have correspondingly large masses. 
}

\FullConference{EPS-HEP 2017, European Physical Society conference on High Energy Physics\\
		5-12 July 2017\\
		Venice, Italy}

\begin{document}

Recently~\cite{Fernandez:2015zsa}, we have proposed a workable reason why the SM groups are so small based on spontaneous mass generation in strongly coupled gauge theories.
Other authors~\cite{Kleppe:2014ula} have also recently pondered why these groups instead of others, and they guess that the smallness of their representations and corresponding Casimir invariants play a role. 
However, in the fundamental representation, the Casimir  $C_F$ invariants grow linearly with the group dimension as shown in table~\ref{tcolor} so that $SU(4)$, $SU(5)$ and groups of equal dimension are not so dissimilar to QCD.
\begin{table}[h]
\centering
\resizebox*{!}{5cm}{
\begin{tabular}{|c||c|}
\hline
Group & Color Factor ($C_F$) \\ \hline
$SU(N_{c})$ & $\frac{1}{2}\Big(N_{c}-\frac{1}{N_{c}}\Big) \ \ \ \forall N_{c}\in \mathbb{N} $ \\  \hline
$SO(N_{c})$ & $\frac{1}{4}\Big(N_{c}-1\Big) \ \ \ \forall N_{c}\in\mathbb{N}$\\ \hline
$Sp(N_{c})$ & $\frac{1}{4}\Big(N_{c}+1\Big) \  \ \ N_{c}=2n \ \ n\in\mathbb{N}$ \\ \hline
$E6$ & $\frac{1}{12}\Big(N_{c}-\frac{29}{3}\Big) \ \ N_{c}=27$ \\ \hline
$F4$ & $\frac{1}{18}\Big(N_{c}-8\Big) \ \ N_{c}=26$ \\ \hline
$G2$ & $\frac{1}{4}\Big(N_{c}-3\Big) \ \ N_{c}=7$ \\ \hline
$E7$ & $\frac{1}{48}\Big(N_{c}+1\Big) \ \ N_{c}=56$ \\ \hline
\end{tabular}}
\caption{Color factors $C_F$ for fermions in the fundamental representation of all classical and most special (simple, compact) Lie groups~\cite{Cvitanovic:1976am,vanRitbergen:1998pn}. The Casimir $C_F$ is seen to grow linearly with the group dimension.
}
\label{tcolor}
\end{table}   

\begin{figure}[h]
\centerline{\includegraphics*[width=8cm]{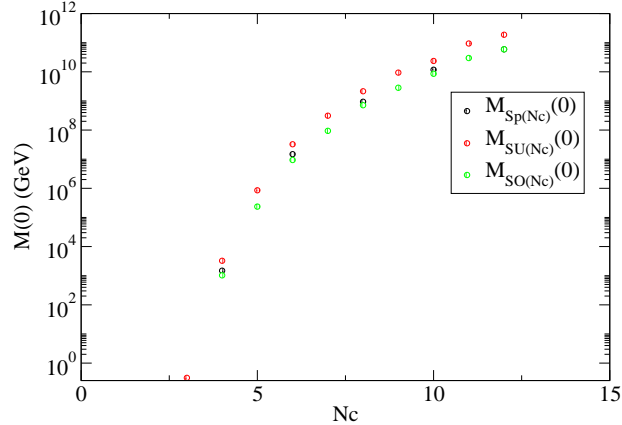}}
\caption{Dynamical mass as function of $N_c$ from 3 to 12  obtained by 
matching perturbation theory and DSE when 
$(C_F\alpha_s)=0.4$, and obtaining the DSE solution by rescaling that of $SU(3)$.
The advertised exponential dependence at low scales is plain. The masses of fermions charged under very large groups then cluster around the starting Grand Unification scale.
\label{fig:mofNc}}
\end{figure}

To achieve a clear separation of larger groups, a selection criterion that is exponential in the number of colors, rather than linear, is preferable. We have found that the mechanism of spontaneous chiral symmetry breaking generates a fermion mass that indeed scales exponentially with the number of colors if all groups are about equally strongly coupled at a large scale of order $10^{15}$--$10^{16}$ GeV. This is shown in figure~\ref{fig:mofNc}.

This exponential dependence removes fermions charged under large groups from the low--energy spectrum. To obtain the result we used the scaling properties of a simplified Momentum-subtracted Dyson-Schwinger equation for the running fermion mass $M(p^2)$ (highlighted in blue; the group--theoretical color factor is also highlighted in red)  
\begin{equation}
\textcolor{blue}{M(p^2)}= M(\mu^2) + \textcolor{red}{C_F} 
\frac{4\alpha}{\pi^2}\int_0^\infty q^3dq
 \int_{-1}^1dx \sqrt{1-x^2}  \nonumber 
\left( \frac{1}{ \left| q  - p \right|^2 }- 
       \frac{1}{ \left| q  -\mu\right|^2 } 
\right)  
       \frac{\textcolor{blue}{M(q^2)}}{\textcolor{blue}{M^2(q^2)}+ \left| q \right|^2 }.
\end{equation}
This allows to establish a proportionality 
\begin{equation}
\frac{M_{group}(0)}{M_{SU(3)}(0)} = \frac{\sigma_{group}}{\sigma_{SU(3)}}
\end{equation}
between induced masses $M(p^2=0)$ and the scales at which the respective groups become strongly coupled due to the running of their coupling constant $\sigma$ down from $\mu$. 

The renormalization scale where both the coupling constant and the fermion masses are fixed 
(to $\alpha=0.017$ and $m\simeq 1$MeV) is $\mu=10^{15}$ GeV. These values are chosen so that the QCD coupling constant at the Z-pole and isospin-averaged light quark mass at 2 GeV take their known values while evolving from $\mu$ with the one--loop renormalization group equation.

\begin{figure}[h]
\centerline{\includegraphics*[width=8cm]{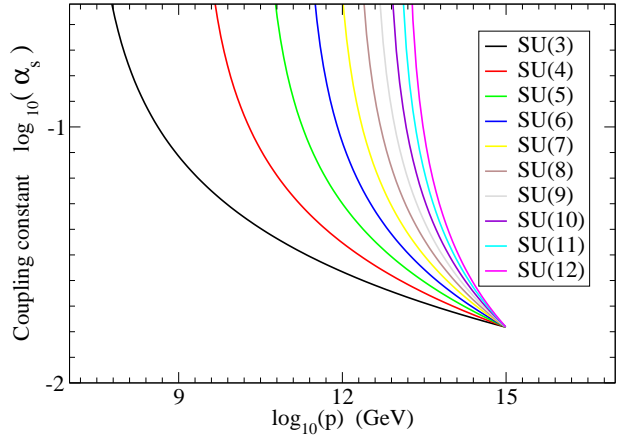}}
\caption{One-loop running coupling for $SU(N_c)$ ($N_c\geq 3$, $N_f=1$). 
For other groups the running is qualitatively similar since it depends, through $\beta_1=\frac{1}{6}(11N_{c}-2N_{f})$, on the group's fundamental dimension only. All couplings are chosen to be the same at the GUT scale $10^{15}$ GeV.
\label{fig:runningalpha}}
\end{figure}

The scale $\sigma$ where the coupling constant becomes strong (which we take to mean
$C_F \alpha_s = 0.4$, which would roughly correspond to the charmonium scale in QCD--the benchmark--)
is reminded in figure~\ref{fig:runningalpha}.

The largest group represented, $SU(12)$, is seen to become strongly coupled just below the GUT scale.

In conclusion, we cannot predict whether there are (or not) fermions charged under larger Lie groups than are seen in the known Standard Model spectrum. But we find a simple negative answer: if there is something to Grand Unification ideas and indeed a new scale of physics around $10^{15}$ GeV emerges, and if those other Lie-based gauge theories have all similar couplings there, then
the fermions charged under those groups have very large masses and are not yet visible.

Thus, the situation could roughly be that the electroweak $U(1)\times SU(2)_L$ model is too weakly coupled to feature spontaneous mass generation (an explicit Higgs field is necessary), QCD is just rightly coupled to trigger the mechanism at the GeV scale, and groups of larger dimension also have spontaneously generated masses, but the scale is exponentially larger~\footnote{Holger Nielsen has also given a temptative reason why the groups of small dimension cannot repeatedly appear: the effect of rotating fermions by several phases from different equal groups could not be distinguished from a phase rotation in a single one of them, so that one does not need to consider copies of the same group. We thank Pedro Bicudo for this comment.}.

A final feature of spontaneous symmetry breaking that we need to discuss is the presence of (quasi)Goldstone bosons. Upon generating a fermion mass and breaking the global symmetry, 
a family of pion-like pseudoscalar particles is necessary, with typical mass below that of the fermions. The Gell-Mann-Oakes-Renner relation $M_\pi^2 f_\pi^2 = -2m_q \langle \bar{q}q\rangle$ suggests that their mass is of order
\begin{equation}
M_\pi(N_c) \sim \sqrt{\sigma(N_c)\ m_q(\sigma(N_c))}\ . 
\end{equation}

What does the nonobservation of such pseudoscalar particles up to energies of order 1 TeV imply? 
Proceeding to larger dimension than in the Standard Model, fermions charged under $SU(4)$ would inhabit the $\sigma(4)\sim 100-1000$ TeV region. With $m_q\sim 1$ MeV, we see that the corresponding Goldstone bosons would have masses
\begin{equation}
M_\pi(SU(4)) \sim 10\ {\rm GeV}
\end{equation} 
which would put them in the bottomonium region. Seeing how hard it was to discover the pseudoscalar $\eta_b$, this possibility should be left open for more careful investigation.

To proceed beyond qualitative statements, one would need to achieve precision calculations in gauge theories with large groups which become strongly coupled. Lattice gauge theory is already being deployed~\cite{Maas:2016ngo} in studies beyond the Standard Model and one could imagine a precise prediction of the masses of $SU(4)$-charged fermions and corresponding Goldstone bosons becoming available.



\begin{thebibliography}{99}

\bibitem{Fernandez:2015zsa}
  G.~Garc\'{\i}a Fern\'andez, J.~Guerrero Rojas and F.~J.~Llanes-Estrada,
  Nucl.\ Phys.\ B {\bf 915} (2017) 262
  doi:10.1016/j.nuclphysb.2016.12.010
  [arXiv:1507.08143 [hep-ph]].


\bibitem{Kleppe:2014ula}
  A.~Kleppe and H.~B.~Nielsen,
  Bled Workshops Phys.\  {\bf 15} (2014) no.2,  267
  [arXiv:1412.7497 [gr-qc]].




 


\bibitem{Cvitanovic:1976am} 
  P.~Cvitanovic,
  Phys.\ Rev.\ D {\bf 14}, 1536 (1976);
P.Cvitanovic: \textit{Group Theory: Birdtracks, Lie's, and exceptional groups}, Princeton University Press, Princeton, New Jersey (2008).

\bibitem{vanRitbergen:1998pn} 
  T.~van Ritbergen, A.~N.~Schellekens and J.~A.~M.~Vermaseren,
  Int.\ J.\ Mod.\ Phys.\ A {\bf 14}, 41 (1999).


\bibitem{Maas:2016ngo} 
  A.~Maas and P.~Torek,
  Phys.\ Rev.\ D {\bf 95}, no. 1, 014501 (2017)
  doi:10.1103/PhysRevD.95.014501
  [arXiv:1607.05860 [hep-lat]].

\end{thebibliography}
\end{document}